\documentclass[
 reprint,superscriptaddress,
nofootinbib,
 amsmath,amssymb,
 aps,prd,
floatfix
]{revtex4-2}

\usepackage[utf8]{inputenc}
\usepackage{color}
\usepackage{graphicx}
\usepackage{dcolumn}
\usepackage{bm}
\usepackage{braket}
\usepackage{comment}
\usepackage[version=3]{mhchem}
\usepackage{soul}
\usepackage{hyperref}
\usepackage{fontawesome5}

\definecolor{deepgreen}{rgb}{0.2,0.8,0.2}

\definecolor{deepblue}{rgb}{0.2,0.2,0.8}

\definecolor{deepred}{rgb}{0.8,0.2,0.2}

\newcommand{\vect}[1]{\boldsymbol{\mathbf{#1}}}

\newcommand{\dd}{{\rm d}}
\newcommand{\cc}{\text{c.c.}}

\usepackage{todonotes}

\begin{document}

\title{Detecting the QCD axion via the ferroaxionic force with piezoelectric materials}
\author{Asimina Arvanitaki}

\email{aarvanitaki@perimeterinstitute.ca}
\affiliation{Perimeter Institute for Theoretical Physics, Waterloo, Ontario N2L 2Y5, Canada}

\author{Jonathan Engel}
\email{engelj@physics.unc.edu}
\affiliation{Department of Physics and Astronomy, University of North Carolina,
Chapel Hill, North Carolina 27516-3255, USA}

\author{Andrew A.~Geraci}
\email{andrew.geraci@northwestern.edu}
\affiliation{Center for Fundamental Physics, Department of Physics and Astronomy, Northwestern University, Evanston, IL 60208}

\author{Alexander Hepburn}
\email{ahepburn@unc.edu}
\affiliation{Department of Physics and Astronomy, University of North Carolina,
Chapel Hill, North Carolina 27516-3255, USA}

\author{Amalia Madden}
\email{amadden@kitp.ucsb.edu}
\affiliation{Kavli Institute for Theoretical Physics, University of California, Santa Barbara, CA 93106, USA}
\affiliation{Perimeter Institute for Theoretical Physics, Waterloo, Ontario N2L 2Y5, Canada}

\author{Ken Van Tilburg}
 \email{kenvt@nyu.edu}
 \email{kvantilburg@flatironinstitute.org}
 \affiliation{Center for Cosmology and Particle Physics, Department of Physics, New York University,
New York, NY 10003, USA}
 \affiliation{Center for Computational Astrophysics, Flatiron Institute, New York, NY 10010, USA}
 
\date{\today}

\begin{abstract}
We show that piezoelectric materials can be used to source virtual QCD axions, generating a new axion-mediated force. Spontaneous parity violation within the piezoelectric crystal combined with time-reversal violation from aligned spins provide the necessary symmetry breaking to produce an effective in-medium scalar coupling of the axion to nucleons up to 7 orders of magnitude larger than that in vacuum. We propose a detection scheme based on nuclear spin precession caused by the axion’s pseudoscalar coupling to nuclear spins. This signal is resonantly enhanced when the distance between the source crystal and the spin sample is modulated at the spin precession frequency.  Using this effect, future experimental setups can be sensitive to the QCD axion in the unexplored mass range from $10^{-5}\,\mathrm{eV}$ to $10^{-2}\,\mathrm{eV}$.
\end{abstract}

\maketitle


\section{Introduction}

Axions, parity-odd spinless bosons, are some of the best-motivated particles beyond the Standard Model of particle physics. The field excitation associated with a mechanism to explain the smallness of the neutron's electric dipole moment (EDM) $d_n$~\cite{Peccei:1977hh, Weinberg:1977ma, Wilczek:1977pj, Dine:1982ah, Preskill:1982cy} (the strong $\mathsf{CP}$ problem) is known as the QCD axion, named for its irreducible coupling to gluons. Weakly-coupled axions are embedded into string theory without contrivance~\cite{Svrcek:2006yi}, and for string compactifications with sufficiently complex topology there can exist a plenitude of light axions~\cite{Arvanitaki:2009fg}. They are also excellent dark matter (DM) candidates that are generically produced in the early universe.

Axion couplings are constrained by their transformations under parity ($\mathsf{P}$), time-reversal ($\mathsf{T}$), and shift symmetry. These approximate symmetries protect the axion's mass from quantum corrections, and imply that the axion primarily couples to matter through derivative axial couplings, and to gauge bosons via $\mathsf{P}$- and $\mathsf{T}$-odd operators. The QCD axion, defined as the linear combination $a$ that couples to gluons (with field strength $G^c_{\mu \nu}$)
\begin{align}
\mathcal{L} \supset \frac{\left(\partial a\right)^2}{2} 
+ \frac{a}{f_a}\frac{\alpha_s}{8\pi} \epsilon^{\mu \nu \rho \sigma} G^c_{\mu \nu} G^c_{\rho \sigma}, \label{eq:aGG}
\end{align}
is special. QCD instanton effects break its shift symmetry non-perturbatively, generating its mass and non-derivative couplings proportional to the inverse of the decay constant $f_a$, such as interactions with the electric dipole moments of nucleons. Scalar $\mathsf{P}$- and $\mathsf{T}$-even couplings to matter can only arise in the presence of $\mathsf{CP}$ violation in the theory, and can generate a monopole-dipole force between unpolarized and polarized matter~\cite{moody1984new}, a phenomenon sought after in a variety of experiments~\cite{Arvanitaki:2014dfa, geraci2017progress}. 

Ref.~\cite{Arvanitaki:2021wjk} pointed out that inside a crystal with a $\mathsf{P}$-violating (piezoelectric) lattice structure and polarized nuclear spins which provide the breaking of $\mathsf{T}$ invariance, a QCD axion DM background can manifest itself as a periodic stress that coherently excites a macroscopic strain via Eq.~\ref{eq:aGG}. We coined this phenomenon ``the (converse) piezoaxionic effect'', and proposed a new class of DM searches for the QCD axion based on bulk acoustic resonators. Previous works~\cite{Graham:2013gfa,Budker:2013hfa} recognized that the same coupling can also cause nuclear spin resonance under similar conditions.
Just as piezoelectricity consists of a direct effect (stress generating an electric field) and a converse effect (an electric field producing stress), piezoaxionicity can also be employed to directly generate an axion field by applying stress.

In this Letter, we discuss a new signature of the QCD axion---the ferroaxionic force---wherein a spin-polarized piezoelectric crystal sources a static monopole axion field through the defining coupling of Eq.~\ref{eq:aGG}, \emph{without} any applied stress. The resulting in-medium scalar coupling can be orders of magnitude larger than its predicted in-vacuum scalar coupling, without the need for additional $\mathsf{CP}$ violation in the fundamental theory. The ferroaxionic force can be used as a model-independent source for monopole-dipole forces mediated by the QCD axion, thus motivating a new class of short-range force experiments.

This paper is structured as follows. In Sec.~\ref{sec:Theory}, we explain how this new coupling arises, and estimate its effective size for some candidate nuclei and crystals. In Sec.~\ref{sec:Setup}, we describe an experimental setup that can probe the ferroaxionic force, based on a modification of the ARIADNE experiment~\cite{Arvanitaki:2014dfa}. The forecasted sensitivity of this setup is discussed in Sec.~\ref{sec:Sensitivity}. We conclude in Sec.~\ref{sec:Conclusions}. We employ units wherein $\hbar = c = 1$. The code for this paper can be found on GitHub (\href{https://github.com/kenvantilburg/ferroaxionic-effect}{%
  \faGithub
}).

\section{Theory}
\label{sec:Theory}

\paragraph*{Axion interactions---}
At low energies below QCD confinement, Eq.~\ref{eq:aGG} reduces to an effective Lagrangian for the QCD axion coupled to nucleons $N$:
\begin{align}
\mathcal{L} 
&\supset \frac{\left(\partial a\right)^2}{2} - \frac{m_a^2 a^2}{2} - \frac{a}{f_a} \left(\rho_\mathsf{S} + \rho_\mathsf{M}\right) \nonumber \\
&\phantom{\supset} - g_{s}^N a \bar{N} N  + \frac{g_{p}^N}{2 m_N} \partial_\mu a \bar{N} \gamma^\mu \gamma_5 N. \label{eq:L}
\end{align}
We ignore couplings to electrons and photons.
The axion mass is inversely related to the decay constant $f_a$ as $m_a \approx \left(5.7 \pm 0.07\right)\,\mathrm{meV} \left({10^{9}\,\mathrm{GeV}}/{f_a}\right)$~\cite{di2016qcd}. The factors $\rho_\mathsf{S}$ and $\rho_\mathsf{M}$ are effective in-medium energy densities from interactions of the nuclear Schiff moments (SM) $\mathsf{S}$ and magnetic quadrupole moments (MQM) $\mathsf{M}$ in the material; they are the subject of this work and will be estimated below.

First, we quote the expected in-vacuum scalar ($g_{s}^N$) and pseudoscalar ($g_{p}^N$) couplings to nucleons. The scalar coupling is $\mathsf{CP}$ violating and proportional to the CKM phase~\cite{Georgi:1986kr}:
\begin{align}
    g_{s}^N &\sim 10^{-30} \, \frac{10^9\,\text{GeV}}{f_a} \label{eq:gsN_minimal}
\end{align}
for both $N = \text{proton},~\text{neutron}$ (isospin breaking is suppressed).
This numerical estimate is highly uncertain but unlikely to much smaller than in Eq.~\ref{eq:gsN_minimal}~\cite{Okawa:2021fto}. Additional sources of $\mathsf{CP}$ violation beyond the Standard Model that shift the minimum of the axion potential to $a / f_a = \overline{\theta}_\text{ind}$ can boost it to $g_{s}^N \approx 1.5 \times 10^{-21} \left({\overline{\theta}_\text{ind}}/{10^{-10}}\right) \left({10^9\,\text{GeV}}/{f_a}\right)$~\cite{Okawa:2021fto}, saturating the experimental upper limit on the neutron EDM~\cite{Abel:2020pzs}.
The pseudoscalar coupling to nucleons is not $\mathsf{CP}$ violating, and has a generic size of:
\begin{align}
    g_{p}^N \equiv \frac{c_N m_N}{f_a} \approx c_N \times 10^{-9} \,\frac{10^9\,\text{GeV}}{f_a}. \label{eq:g_p}
\end{align}
The interaction of Eq.~\ref{eq:aGG} irreducibly yields the coefficients $c_{\text{proton}} \approx 0.47(3)$ and $c_{\text{neutron}} \approx 0.02(3)$, e.g.~in the KSVZ model~\cite{Kim:1979if,Shifman:1979if}. These can receive corrections from derivative couplings to quarks: in DFSZ axion models~\cite{Dine:1981rt,Zhitnitsky:1980tq},  $c_{\text{proton}} \approx -0.617 + 0.435 \sin^2 \beta \pm 0.025$ and $c_{\text{neutron}} \approx 0.254 - 0.414 \sin^2 \beta \pm 0.025$~\cite{di2016qcd}.

\paragraph*{Axion sourcing---}
A $\mathsf{P}$- and $\mathsf{T}$-odd ``monopole'' axion configuration is generated via the equation of motion:
\begin{align}
    \left(\Box + m_a^2\right) a\left(t,\vect{x}\right) = -\frac{\rho_\mathsf{S} + \rho_\mathsf{M}}{f_a} - g_{s}^N n_N  \equiv j(t,\vect{x}). \label{eq:EOM}
\end{align}
The source of $\mathsf{P}$ and $\mathsf{T}$ violation on the RHS can be \emph{explicit} through $g_{s}^N$ and/or \emph{spontaneous} through the in-medium SM or MQM energy densities $\rho_\mathsf{S}$ and $\rho_\mathsf{M}$; we shall see that the latter two can generate larger axion fields than the scalar coupling of Eq.~\ref{eq:gsN_minimal}. In the quasistatic approximation, the axion field solution to Eq.~\ref{eq:EOM} is 
$a(t,\vect{x}) = \int \dd^3 x' \, (4\pi |\vect{x} - \vect{x}'|)^{-1} e^{-m_a |\vect{x} - \vect{x}'|} j(t,\vect{x}')$.
A distance $\vect{D}$ away from a uniform slab of thickness $h$ and large transverse area $A \gg h^2, D^2$, the gradient of the source-induced $\overline{\theta}_a = a /f_a$ angle is approximately:
\begin{align}
    \vect{\nabla} \overline{\theta}_a \simeq - \hat{\vect{D}} \frac{j}{2 m_a f_a} e^{-m_a D} \left(1 - e^{-m_a h}\right), \label{eq:grad_theta_mon}
\end{align}
where (slow) time dependence may enter implicitly either through $\vect{D}$ or $j$.

Similarly, a ``dipole'' axion field configuration may be sourced from an ensemble of spins $\vect{\sigma}_N$ with number density $n_N$ via the pseudoscalar coupling~\cite{moody1984new, Arvanitaki:2014dfa}:
$a(t,\vect{x}) = (g_p^N/8\pi m_N) \int \dd^3 x' \, n_N(\vect{x}') e^{-m_a|\vect{x}-\vect{x}'|} |\vect{x}-\vect{x}'|^{-3} (1 + m  |\vect{x}-\vect{x}'|)  (\vect{x}-\vect{x}') \cdot \vect{\sigma}_N$, which yields the $\overline{\theta}_a$ angle gradient
\begin{align}
    \vect{\nabla} \overline{\theta}_a \simeq - \hat{\vect{D}} \left(\hat{\vect{D}} \cdot \vect{\sigma}_N\right) \frac{g_p^N n_N}{4 m_N f_a} e^{-m_a D} \left(1-e^{-m_a h}\right). \label{eq:grad_theta_dip}
\end{align}
from a uniform slab of large transverse area.
The parametric ratio of Eqs.~\ref{eq:grad_theta_mon} \&~\ref{eq:grad_theta_dip} is $2 j / c_N m_a f_a n_N$ after using Eq.~\ref{eq:g_p}.

The gradient of the induced theta angle is the observable quantity of interest, as detection is most sensitively performed via the Hamiltonian
\begin{align}
    H \supset - \frac{g_p^N}{m_N} \vect{\sigma}_N \cdot \vect{\nabla} \overline{\theta}_a \equiv - \Omega_\theta \, \hat{\vect{D}} \cdot \vect{\sigma}_N, \label{eq:H_det}
\end{align}
through a spin precession process analogous to nuclear magnetic resonance (NMR), with Rabi frequency $\Omega_\theta$. The combination of Eqs.~\ref{eq:grad_theta_mon} \&~\ref{eq:H_det} (Eqs.~\ref{eq:grad_theta_dip} \&~\ref{eq:H_det}) yields a monopole-dipole (dipole-dipole) potential. Ref.~\cite{Arvanitaki:2014dfa} has charted out the sensitivity of these potentials proportional to $g_s^N g_p^N$ and $(g_p^N)^2$, respectively. In this work, we propose sourcing a QCD axion gradient through its \emph{irreducible} coupling, via the spontaneous $\mathsf{P}$ and $\mathsf{T}$ violation---parametrized by the energy densities $\rho_\mathsf{S}$ and $\rho_\mathsf{M}$ of SMs and MQMs---that a parity-odd crystal with polarized nuclear spins exhibits.

\paragraph*{Schiff Moment---}
The QCD axion field generated via the SM is proportional to:
\begin{align}
    \rho_\mathsf{S} 
    &= 4 \pi e n_\mathrm{S} \frac{\partial \mathsf{S}}{\partial \overline{\theta}_a} \bm{\mathcal{M}}_\mathsf{S} \cdot \hat{\vect{I}} \label{eq:rho_S} \\
    \vect{\mathcal{M}}_\mathsf{S} 
    &= \sum_{j,m_j, m_{j'}}  \epsilon^*_{p_j,m_j} \epsilon_{s, m_{j'}} \vect{\mathcal{M}}_{\mathsf{S}:\,j,m_j, m_{j'}}+ \text{c.c.}
    \label{eq:M_S},
\end{align}
where $n_\mathsf{S}$ is the density of nuclei with large Schiff moments, and $\vect{I}$ is the nuclear spin vector, normalized such that $|\hat{\vect{I}}| = 1$ corresponds to a fully polarized nuclear spin state. 
The atomic matrix element~\cite{Khriplovich:1997ga,Arvanitaki:2021wjk}
\begin{equation}\bm{\mathcal{M}}_{\mathsf{S}:\,j,m_j, m_{j'}} =  \braket{\Omega_{p_j, m_j}\lvert {\hat{\vect{r}}}\rvert \Omega_{s, m_{j'}}} \frac{Z^2}{a_0^4(\nu_s\nu_{p_j})^{3/2}}\mathcal{R}_j,
\label{eq:renormalisedM}
\end{equation}
quantifies the mixing of atomic orbitals by the nuclear SM potential, and has a
direction given by the intrinsic electric polarization vector of the (necessarily pyroelectric) crystal. In Eq.~\ref{eq:renormalisedM}, $\Omega_i$ are spherical spinor wavefunctions~\cite[\S35]{Berestetskii:1982qgu} and $\mathcal{R}_j$ is a relativistic enhancement factor~\cite{Khriplovich:1997ga} as large as $\sim 9$ for large nuclei.
The quantum numbers $j$ and $m_j$ are those of the total angular momentum and its projection on the $z$-axis, respectively. The coefficients $\epsilon_{s, m_{j'}}$ and $\epsilon_{p_j, m_j}$ parametrize the admixture of atomic $s$ and $p$ valence electron states, i.e.~$\ket{\psi_{\text{el}}} = \sum_{m_{j'}} \epsilon_{s,m_{j'}}\ket{s^0_{m_{j'}}} + \sum_{j, m_j} \epsilon_{p_j, m_j} \ket{p^0_{j, m_j}}$, and characterize the breaking of parity symmetry by the crystal potential.

The axion signal from the SM is largest for heavy, deformed nuclei with a large SM dependence on $\overline{\theta}_a$ and a large number density $n_\mathsf{S}$ in a suitable crystal. The ferroaxionic force is proportional to the dimensionless vector $\epsilon_p^*\epsilon_s \braket{\Omega_p\lvert {\hat{\vect{r}}}\rvert \Omega_{s}}$, whose $\mathcal{O}(1)$ matrix elements are listed in Ref.~\cite[App.~A]{Arvanitaki:2021wjk}, and can only have nonzero expectation value in the pyroelectric crystal classes~\cite{nye1985physical}. Based on naive dimensional analysis, the wavefunction admixtures $\epsilon_{s,p}$ should also be $\mathcal{O}(1)$ in strong pyroelectrics, but a precise determination requires input from ab initio methods such as density functional theory (DFT)~\cite{Skripnikov2016}. 

\paragraph*{Magnetic Quadrupole Moment---} 
The effective MQM energy density in Eq.~\ref{eq:EOM} is given by \cite{Khriplovich:1997ga}:
\begin{align}
    \rho_\mathsf{M} &= e n_\mathsf{M} \frac{\partial \mathsf{M}}{\partial \overline{\theta}_a}\, t_{fg} A_{fg}
    \label{eq:rho_M}
    \\
    t_{fg} &= \frac{1}{4 I \left(2I-1\right)}\left[I_f I_g + I_g I_f -\frac{2}{3} \delta_{fg} I(I+1) \right] 
    \\
    A_{fg} &= \frac{Z^2 \alpha}{a_0^3 (\nu_s\nu_{p_{3/2}})^{3/2}} \,\mathcal{C}
    \sum_{m_j, m_{j'}}\epsilon^*_{p_{3/2},m_j}\epsilon_{s, m_{j'}} 
    \label{eq:Amk} \\
    &\braket{\Omega_{p_{3/2}, m_j}\lvert { \sigma_f \hat{r}_g + \sigma_g \hat{r}_f 
    - 2(\bm{\sigma}\cdot \bm{\hat{r}}) \hat{r}_g \hat{r}_f }\rvert \Omega_{s, m_{j'}}} + \cc \nonumber
\end{align}
where $\mathsf{M}$ is the magnitude of the nuclear MQM, $n_\mathsf{M}$ is the number density of nuclei with an MQM, $\sigma$ is the electron angular momentum operator, $\mathcal{C}$ is an $\mathcal{O}(1)$ numerical coefficient detailed in App.~\ref{app:mqm_elements}, and the the remaining symbols are the same as those defined for the Schiff moment. Only nuclei with spin $I \geq 1$ may exhibit an MQM. Since the MQM operator is an irreducible second-rank tensor in electron variables, it does not mix $p_{1/2}$ and $s_{1/2}$ states~\cite{Khriplovich:1997ga}. 

Since the MQM operator is linear in electron spin, obtaining $\rho_\mathsf{M} \neq 0$ requires magnetic ordering. The first two terms in the second line of Eq.~\ref{eq:Amk} are linear in the radial vector and thus non-zero in a pyroelectric material due to its spontaneous electric polarization vector. The last, cubic term is a parity-odd, rank-three tensor in electron position space, with symmetry properties shared by the material's piezoelectric tensor. Therefore, non-zero $\rho_\mathsf{M}$ can also be generated in the broader class of piezoelectric crystals, of which pyroelectrics are a subset. We list individual MQM matrix elements in App.~\ref{app:mqm_elements}. Like for the SM effects, we expect $\epsilon_{p}^* \epsilon_s$ to be commensurate with the strength of the piezoelectric/pyroelectric tensors, and the second line of Eq.~\ref{eq:Amk} to be $\mathcal{O}(1)$ numerically.

\begin{table}[!htbp]
    \centering
\bgroup
\def\arraystretch{1.25}
\setlength{\tabcolsep}{0.8em} 
\begin{tabular}{l l l l l}
    \hline \hline 
    & $\mathsf{S}$~$\left[\,\overline{\theta}_a\, e \,\text{fm}^3\right]$ & Ref. &
    $\mathsf{M}$~$\left[\,\overline{\theta}_a\, e \,\text{fm}^2\right]$ \\
    \hline
    \ce{^{153}_{63}Eu} & 0.15   & \cite{Sushkov:2023myw}    & 1.0 \\
    \ce{^{235}_{92}U}  & 3      & \cite{Flambaum:2019tym}   & 3.1 \\
    \ce{^{237}_{93}Np} & 0.75   & \cite{Sushkov:2023myw}    & 1.3 \\
    \hline \hline
    \end{tabular}
\egroup
    \caption{Schiff moments $\mathsf{S}$ and magnetic quadrupole  moments $\mathsf{M}$ of deformed nuclei. The SMs are taken from the listed references, while the MQMs are estimated using the results of Refs.~\cite[Ch.~10]{Khriplovich:1997ga} \&~\cite{Engel:1999np} with pion-nucleon couplings from Ref.~\cite{deVries:2020iea}, similar to the approach of Ref.~\cite{Arvanitaki:2021wjk}. 
    }
    \label{tab:materials}
\end{table}
The ideal nucleus is stable or meta-stable and has a large SM and/or MQM, which can be enhanced in nuclei with intrinsic octupole or quadrupole deformation, respectively~\cite{Flambaum:1984fb, Khriplovich:1997ga, Engel:1999np}. We suggest suitable nuclei in Tab.\ \ref{tab:materials} with values of their SM and MQM taken from the literature. Some may be optimistic.  We have carried out computations of the Schiff moment of $^{153}$Eu, in the same nuclear-density-functional-based approach as was used to study $^{199}$Hg and $^{211}$Ra in Ref.\ \cite{PhysRevC.82.015501}, with two different Skyrme functionals, and found that in the limit that pions are heavy, we obtain the same result as that of Ref.\ \cite{Sushkov:2023myw} to within a factor of two, but that the finite range interaction associated with the physical pion mass reduces the moments by about a factor of five.  Computations of MQMs are in progress. 

From Eqs.~\ref{eq:gsN_minimal} \&~\ref{eq:EOM}, we can compare the strength of the effective in-medium scalar coupling of the axion, given by the ratio $(\rho_\mathsf{S}+\rho_\mathsf{M})/
{f_a n_N}$, to the vacuum scalar coupling $g_s^N$. Assuming the heavy-nuclei density $n_{\mathsf{S},\mathsf{M}}$ is an $\mathcal{O}(1)$ fraction of the total number density $n_N$ of nuclei, this can reach up to 7 orders of magnitude for the values given in Tab.~\ref{tab:materials}.

\section{Setup}
\label{sec:Setup}

We envision an experimental setup analogous to that
of the ARIADNE experiment described in Refs.~\cite{Arvanitaki:2014dfa,geraci2017progress} with appropriate modifications, including the implementation of a
dilution fridge and a different source mass composition and geometry.
An oblate spheroidal cavity filled with laser-polarized $\ce{^3He}$ can be used to sense the axion gradient generated by the source mass.
The spheroidal shape results in a uniform magnetic field inside the sample when polarized along one of the principal axes~\cite{fosbinder2022method}. 
The source mass generates an axion potential as in Eqs.~\ref{eq:grad_theta_mon} and~\ref{eq:grad_theta_dip} from three distinct mechanisms: (1) the SM density $\propto \rho_\mathsf{S}$, (2) the MQM density $\propto \rho_\mathsf{M}$, and (3) the dipole potential $\propto g_p^N n_N$. All three may be simultaneously generated by the same source mass, and detected through the spin-precession induced by the Hamiltonian of Eq.~\ref{eq:H_det}.

\begin{figure}
    \centering
    \includegraphics[width=1.0\columnwidth]{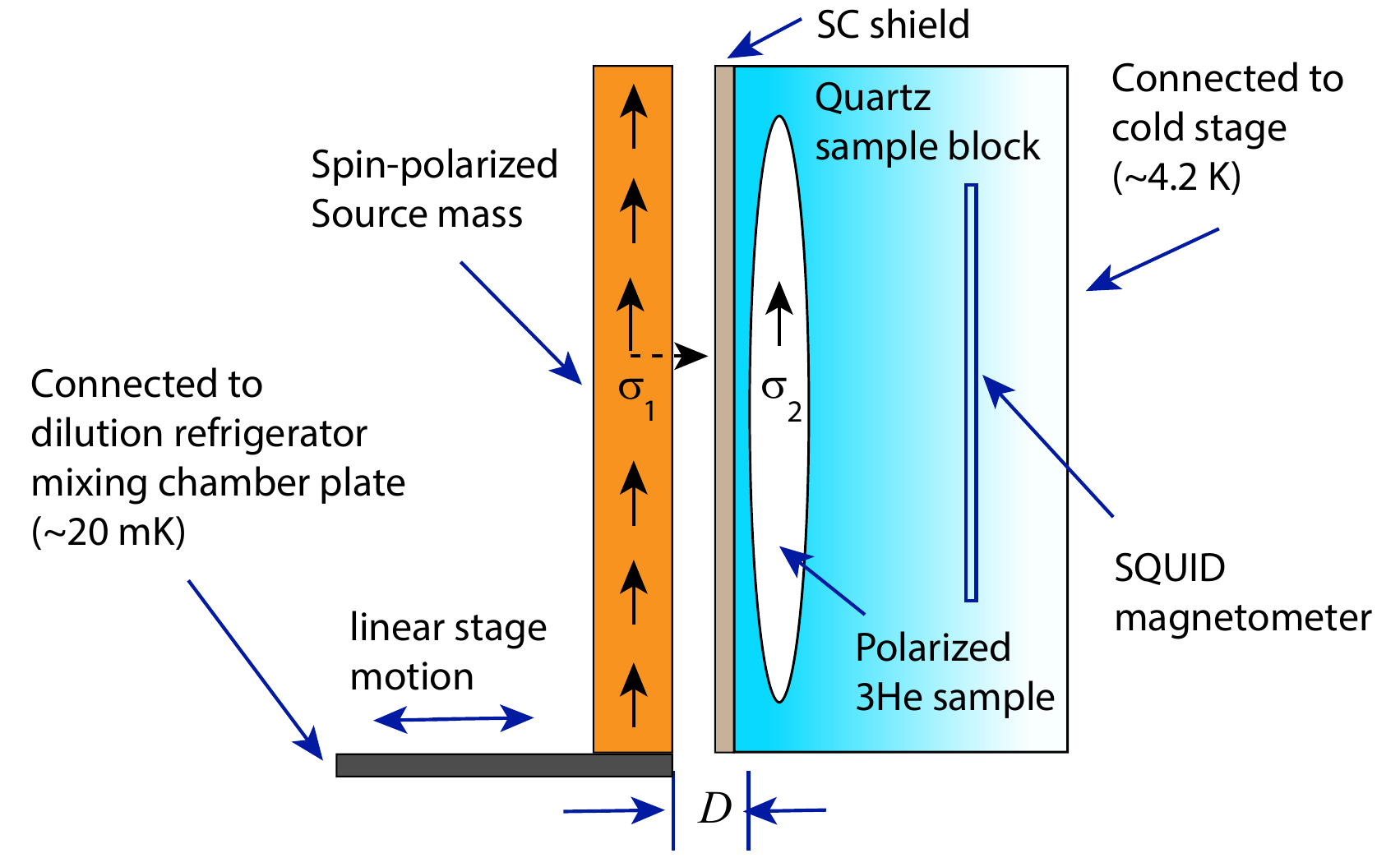}
    \caption{Experimental setup. A polarized \ce{^3 He} sample in an ellipsoidal cavity within a quartz block senses the axion field gradient sourced by a spin-polarized source, which generates a monopole potential from the non-derivative axion coupling to the nuclear SM and MQM, and a dipole potential from the axion's derivative coupling to the nuclear spins. The black arrows indicate the direction of the crystal's pyroelectric vector. The induced transverse magnetization of the precessing \ce{^3 He} spins is read out by a SQUID magnetometer.}
    \label{fig:setup}
\end{figure}

The source mass has a flat-plate geometry to maximize the amount of material near the NMR sample within the Compton wavelength of the axion. For the SM and MQM, the nuclear spins of the source mass should be polarized along an appropriate direction determined by the crystal structure (e.g.~the pyroelectric vector), whereas the dipole-dipole potential is maximized for out-of-plane nuclear spin polarization, as evident from Eqs.~\ref{eq:grad_theta_mon}~\&~\ref{eq:grad_theta_dip} and illustrated in Fig.~\ref{fig:setup}. 

The ellipsoidal \ce{^3 He} sample is inside of a quartz block coated with a layer of superconducting thin-film magnetic shielding and layers of superconducting foils~\cite{fosbinder2022method}. Further details on the magnetic shielding requirements are provided in App.~\ref{app:noise}. The distance $D$ between the source mass and the sample is modulated at the nuclear Larmor frequency to drive coherent Rabi oscillations. For example, a non-magnetic low-temperature linear stage rated down to $\sim 20$ mK temperatures ~\cite{JPE} could actuate the position of the source mass with an amplitude of order $\pm D/2$ for the setups we consider at frequencies ranging from $1-6$ Hz, corresponding to a stage speed of approximately $3$ mm/s. The precessing transverse magnetization is read out using a SQUID magnetometer. 

\paragraph*{Source Materials---}
The required material for this setup is a pyro- or piezo-electric crystal containing a high density of nuclei with large SMs and/or MQMs. Ideally, the material is magnetic so that the aligned electron spins transfer their polarization onto the nuclear spins in thermal equilibrium at a manageable temperature. Potential crystals including the nuclear isotopes of Tab.~\ref{tab:materials} are \ce{LiUO_3}~\cite{kovba:1960}, europium barium titanate \ce{Eu_{0.5}Ba_{0.5}TiO_3}~\cite{10.1063/1.324974}, \ce{Np_3 O F_{12}}~\cite{cousson1985nouvelle}, and \ce{Np I O_5}~\cite{Jain2013}.

\paragraph*{Proposed measurement protocol---}
The experiment is to be performed inside a dilution refrigerator with a large bore and vector magnet, to facilitate polarizing the source mass either in-plane or out-of-plane. The nuclear spins can be polarized using dynamical nuclear polarization, involving electron angular momentum transfer onto nuclei via hyperfine interactions, as detailed in App.~\ref{app:hyperfine}. 
The \ce{^3 He} sample block is maintained at $4\,\mathrm{K}$, while the source mass and its cryogenic linear translation stage~\cite{JPE} connects to the dilution fridge mixing chamber plate and is cooled to $20\,\mathrm{mK}$. After magnetization, the field can be ramped down, and the \ce{Nb} coating of the \ce{^3 He} sample cell heated above $T_\mathrm{c}$ and re-cooled in near-zero field using magnetic shim coils to minimize trapped magnetic flux. Laser-polarized \ce{^3 He} is produced at room temperature above the cryostat using metastability-exchange optical pumping~\cite{MEOP1} before insertion into the sample block for the measurements. 
For optimal sensitivity, the spins of the heavy nuclei in the source crystal should remain polarized at least as long as the transverse spin relaxation time $T_2$ of the polarized \ce{^3 He} detector sample (App.~\ref{app:hyperfine}).

\section{Sensitivity}
\label{sec:Sensitivity}


We consider two concrete geometries, parametrized by the minimum separation $D = 150 \, \mathrm{\mu m}$ and $D= 1 \, \mathrm{mm}$ between the faces of the source crystal and the measurement sample. We assume the source plate has transverse area of $A = (300\,D)^2$ and thickness $h = 5\,D$. We consider the transverse area of the \ce{^3 He} sample to be the same as that of the source crystal.

We expect the sensitivity to be fundamentally limited by quantum spin projection noise in the NMR sample. The magnetic field sensitivity is:
\begin{align}
\delta B &= p^{-1}\sqrt{\frac{2 b}{n_N  \mu_{\ce{^3 He}} \gamma V T_2}} \approx 3 \times 10^{-19} \, {\rm{T}}   \label{eq:B_min} \\
&\times\left( \frac{1}{p} \right) \sqrt{ \left(\frac{b}{1\,\rm{Hz}}\right) \left( \frac{1\,\mbox{mm}^3}{V}\right) \left( \frac{10^{21}\,\mbox{cm}^{-3}}{n_N}\right) \left( \frac{1000\,\rm{s}}{T_2}\right)} . \nonumber
\end{align}
Here, $p$ is the spin polarization fraction, $V$ the sample volume, $\gamma = (2\pi) \times 32.4\,\mathrm{MHz/T}$ is the gyromagnetic ratio for \ce{^3 He}, $b$ is the measurement bandwidth, and $\mu_{^3\rm{He}} = -2.12 \mu_n$ is the \ce{^3 He} nuclear moment \cite{helium_moment} with $\mu_n$ the nuclear magneton.
This corresponds to a Rabi frequency sensitivity
\begin{equation}
    \delta \Omega_\theta \geq \frac{\gamma}{2} \delta B \sim 3 \times 10^{-11} \, \mathrm{rad \, s^{-1}}
\end{equation}
for the same fiducial parameters as in Eq.~\ref{eq:B_min}.

\begin{figure}
\centering
\includegraphics[width=1.0\columnwidth]{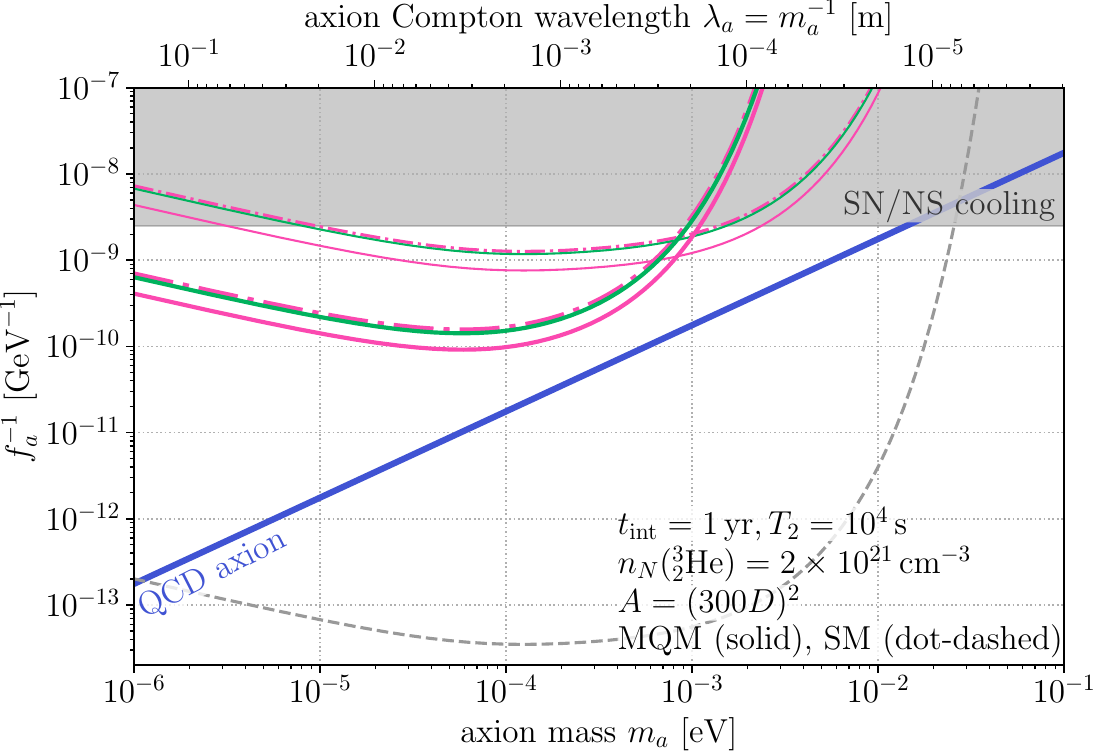}
\caption{Sensitivity to the gluon coupling for a monopole-dipole force generated by the nuclear SM (dot-dashed) or MQM (solid). 
The pink (green) lines correspond to spin-projection-noise limited sensitivity for the crystal \ce{NpIO_{5}} (\ce{Eu_{0.5}Ba_{0.5}TiO_3}). 
Thick (thin) lines corresponds to a source-sample separation $D=1 \, \mathrm{mm}$ ($0.15\,\mathrm{mm}$). The SM and MQM sensitivity curves assume a dipole coupling coefficient $c_N\approx 1.$  
The gray dashed line shows the SQUID-limited sensitivity for the MQM of  \ce{Eu_{0.5}Ba_{0.5}TiO_3}. Supernova (SN) and neutron-star (NS) cooling limits $f_a \gtrsim 4 \times 10^{8}\,\mathrm{GeV}$ for a KSVZ axion~\cite{caputo2024astrophysical}, shown as the gray region.
}
\label{fig:sm-dip}
\end{figure}

\begin{figure}
    \centering
    \includegraphics[width=1.0\columnwidth]{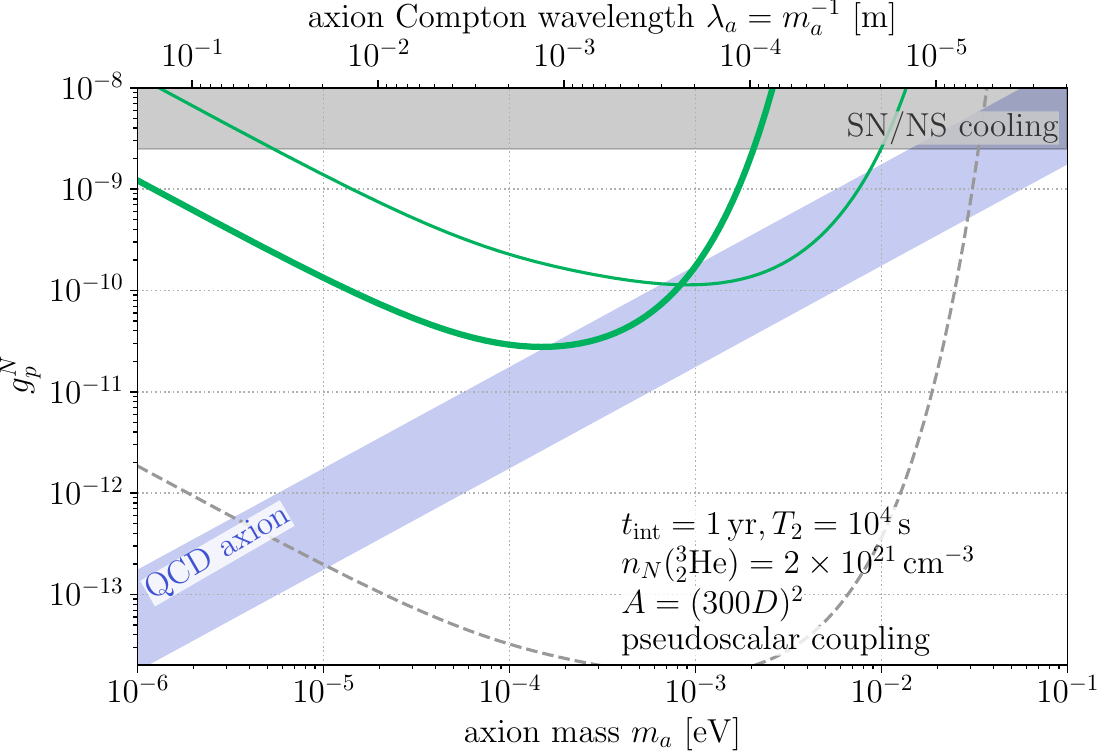}
    \caption{Sensitivity to the pseudoscalar coupling to nucleons for a dipole-dipole force as a function of axion mass $m_a$. The green and dashed gray lines correspond to the same Eu$_{0.5}$Ba$_{0.5}$TiO$_3$ setup as in Fig.~\ref{fig:sm-dip}. The QCD axion band is shown for the range $1 \geq c_N \geq 0.1$.}
    \label{fig:dip-dip}
\end{figure}

Figures~\ref{fig:sm-dip} and~\ref{fig:dip-dip} show the expected sensitivity for a \ce{^3 He} sample with $T_2=10^{4}\,\mathrm{s}$, a spin density $n_N=2 \times 10^{21}\,{\mathrm{cm}}^{-3}$, a distance modulation from $D$ to $2D$, and integration time $t_\mathrm{int} = 1 \, \mathrm{yr}$.  We have chosen to present the reach for the MQM and SM effects separately to emphasize that they correspond to experimentally distinguishable channels, each of which can be isolated with suitable material choices.
We assume a SQUID pickup loop diameter of $5\,\mathrm{cm}$ positioned at a distance of $5D$ from the sample, with a flux sensitivity of $1\,\mu\Phi_0 / \sqrt{\text{Hz}}$. 
The gray dashed line in Figs.~\ref{fig:sm-dip} and~\ref{fig:dip-dip} represents the SQUID-limited sensitivity, which lies well below the QCD axion parameter space. Back-action noise due to the interaction between the spins and the SQUID circuit can also be suppressed~\cite{https://open.bu.edu/handle/2144/46365}. Our current forecasts are limited by spin projection noise at the standard quantum limit.  Spin entanglement in the \ce{^3 He} sample could overcome this limitation by squeezing spin projection noise in the (measured) transverse direction at the expense of increasing the spin noise in an (unmeasured) orthogonal direction~\cite{https://open.bu.edu/handle/2144/46365, PhysRevX.9.021023, PhysRevD.23.1693}. We discuss potential systematic and reducible noise sources in App.~\ref{app:noise}.

\section{Conclusions}
\label{sec:Conclusions}

In this work, we have established a new monopole-dipole force mediated by the QCD axion, generated by parity-violating, magnetically-ordered crystals. This new force could lead to an observable signal in an NMR-like experiment, probing swathes of parameter space complementary to those accessible to cavity experiments and astrophysical phenomena. The origin of this force is a new effective scalar coupling of the QCD axion that derives from its irreducible, $\mathsf{P}$ and $\mathsf{T}$ violating coupling to gluons, together with the large intrinsic $\mathsf{P}$ violation provided by the lattice structure of the piezoelectric crystal, and the $\mathsf{T}$ violation provided by its magnetic ordering. The proposed experimental setup could leverage methods from the ARIADNE experiment currently under construction~\cite{geraci2017progress,ARIADNE:2020wwm}, including techniques for the $^3$He cell and valve construction, $^3$He polarization, magnetic shielding, vibration isolation, and SQUID readout. 
However the rotary stage  would be replaced with a cryogenic-compatible linear stage and thermally connected to the mixing chamber plate of a dilution fridge.

Due to screening of EDMs, the new scalar coupling presented in this work requires either a nuclear SM or MQM. More investigation is needed to confirm the ideal material, as well as DFT calculations to precisely determine the size of atomic parity violation within the crystal. The large difference between the magnetization noise in the NMR sample and the fundamental SQUID noise limit indicates the potential for large sensitivity improvements via spin squeezing procedures.

The setup in this paper will also be sensitive to a dipole-dipole force that could be produced by either the QCD axion or an axion-like particle (Fig.~\ref{fig:dip-dip}). Depending on the size of the model-dependent dipole coupling coefficient $c_N$, this force can be either more or less sensitive than the ferroaxionic monopole-dipole force. Nevertheless, the latter is orders of magnitude larger than the monopole-dipole force  sourced by the $\mathsf{CP}$ violation in the Standard Model in vacuum. Ultimately, only the ferroaxionic force would provide conclusive evidence that the mediating particle is indeed the QCD axion that solves the strong $\mathsf{CP}$ problem.

\acknowledgments{We thank David Stilwell for nuclear structure calculations at the early stages of this work, and Liujun Zou and Alex Sushkov for helpful discussions. AA is grateful for the support of the Stavros Niarchos Foundation and the Gordon and Betty Moore Foundation. 
Research at Perimeter Institute is supported in part by the Government of Canada through the Department of Innovation, Science and Economic Development Canada and by the Province of Ontario through the Ministry of Colleges and Universities. 
JE is supported in part by DOE grant DE-FG02-
97ER41019 and by NSF grant PHY-2110405, which supported the work of AH.
AG is supported in part by NSF grants PHY-2110524 and PHY-2111544, the Heising-Simons Foundation, the W.M. Keck Foundation, the Gordon and Betty Moore Foundation, the Alfred P. Sloan Foundation, the John Templeton Foundation, DARPA, and ONR Grant N00014-18-1-2370. 
AM is supported by NSF grant~PHY-2309135 to the Kavli Institute for Theoretical Physics (KITP) and grant 7392 from the Moore Foundation.
KVT is supported in part by the NSF grant~PHY-2210551. The Flatiron Institute is a division of the Simons Foundation. 
}

\bibliography{ferroaxionic-effect}

\appendix

\section{Nuclear spin polarization}
\label{app:hyperfine}

Without hyperfine interactions, it is exceedingly difficult to obtain near-maximal nuclear spin polarization. In an external magnetic field $\vect{B}_0$ and in thermal equilibrium at a temperature $T$, one has a fractional polarization of: 
\begin{align}
    \langle \hat{\vect{I}} \rangle = \hat{\vect{B}}_0 \tanh\left[\frac{\mu B_0}{2 k_\mathrm{B} T}\right],
\end{align}
with $k_\mathrm{B}$ the  Boltzmann constant.
For $\mu = \mu_N$ (nuclear magneton), $B_0 = 10\,\mathrm{T}$, and $T = 10\,\mathrm{mK}$, we have only $\mu B_0 / 2 k_\mathrm{B}T \approx 0.183$ and thus an equilibrium polarization of just $18\%$ even at these aggressive parameters.

In a magnetic material with polarizable electron spins, the effective magnetic field at the nucleus may be enhanced by hyperfine interactions, allowing for $\mathcal{O}(1)$ nuclear polarization at higher temperatures or in a smaller applied magnetic field. The energy splitting arises from interactions between the nuclear magnetic dipole moment and the electronic magnetic field. Its value is given by~\cite{woodgate1980elementary}:
\begin{align}
    &\hspace{-2em} \braket{\alpha\, J\,I\,F\,M \lvert H_\mathrm{hf} \rvert \alpha'\, J\,I\,F\,M}  = \frac{1}{2}A\,K \nonumber \\
    K &\equiv F(F+1)-J(J+1)-I(I+1)
\end{align}
where $\vect{J} = \vect{L} + \vect{S}$ is electron angular momentum, $\vect{F} = \vect{I} + \vect{J}$ is the total angular momentum of the atom, $M$ is the magnetic quantum number of $F$, $A$ is the magnetic hyperfine structure constant and $\gamma$ specifies the electronic configuration. The hyperfine energy splitting can be related to the effective internal magnetic field $\vect{B}_{\mathrm{int}}$ at the nucleus through $\braket{H_{\mathrm{hf}}} = -\boldsymbol{\mu}\cdot \vect{B}_{\mathrm{int}}$, where $\boldsymbol{\mu}$ corresponds to the nuclear magnetic dipole moment of the specific level. The hyperfine energy splitting also receives a contribution from the interaction between the nuclear electric quadrupole moment with the electronic electric field gradient \cite{woodgate1980elementary}, but for the nuclei listed here this effect is subdominant.

Both of the isotopes $\ce{^{153} Eu}$ and $\ce{^{237} Np}$ have nuclear spins of $I=5/2$ and hyperfine energy splittings of the order of $\Delta E_\mathrm{hf} \sim 10^{-5} \, \text{eV}$ \cite{zemylanoi, stone1968}. For $\ce{^{153} Eu}$, we are considering the $\,^6P_{5/2}$ state, which we assume to be part of the ground-state admixture in a piezoelectric crystal. This energy splitting corresponds to a temperature of around $0.1\,\mathrm{K}$ to achieve $\mathcal{O}(1)$ nuclear spin polarization. 

For $\ce{^{237} Np}$, this higher temperature has the added advantage that it is around the threshold temperature where the radiation emitted by the source can be absorbed by the cooling system. A dilution refrigerator's cooling power  depends on the target temperature $T$ and the $\ce{^3 He}$ flow rate $\dot{n}$~\cite{UHLIG1997279}:
\begin{equation}
\dot{Q}\approx 8.4\,\mathrm{\mu W}~\left(\frac{\dot{n}}{10^{-3}\,\mathrm{mol/s}}\right)\left(\frac{T}{10\,\mathrm{mK}}\right)^2.
\end{equation}
The radiative decay heating power of \ce{^{237} Np} is $0.02\,\mathrm{W/kg}$. For the maximum volume of material considered in this work, $(300 \, \mathrm{mm})^2\times0.5\,\mathrm{cm}$, this limits the crystals containing \ce{^{237} Np}, with densities of around $3\,\mathrm{g/cm^3}$ of Np, to a minimum temperature of around $\sim 0.5\,\mathrm{K}$, which should be sufficient for $\mathcal{O}(1)$ nuclear spin polarization. Another possible meta-stable nuclear candidate that was listed in Tab. \ref{tab:materials}, \ce{^{235} U}, has a lower radioactive heating power of approximately $60 \,\mu \mathrm{W/kg}$.

To achieve optimal sensitivity, the spins of the \ce{^{235} U}, \ce{^{237} Np}, or $\ce{^{153} Eu}$ nuclei in the host material should retain near-unity polarization throughout the measurement time.
This can be achieved by keeping the electron spins aligned \emph{during} the measurement, and thus also the nuclear spins due to the hyperfine interactions. Alternatively, the source crystal can also be demagnetized \emph{before} the measurement (to ease the magnetic shielding requirements), as long as the longitudinal nuclear spin relaxation time $T_1$ is longer than the transverse spin relaxation time $T_2$ of the \ce{^3 He} of the detection sample, and then re-polarized after $T_1$.


\section{Systematic and Magnetic Noise}
\label{app:noise}
In this Appendix, we discuss the systematic noise requirements on the setup.
Several of the systematic noise sources in the envisaged setup are common to those discussed in Refs.~\cite{Arvanitaki:2014dfa, geraci2017progress}, except with the additional complication of a magnetized source crystal. The main systematics are expected to come from magnetic field gradients and background vibrations, magnetic fields that are not adequately screened by the superconducting shield, and noise due to trapped flux in the superconducting shield.

The He sample block is assumed to remain at $T=4$ K so that the $^3$He can remain in a gaseous state to avoid nonlinearities that are known to be challenging for NMR with liquid $^3$He \cite{Hayden:2007} or other hyperpolarized gases such as xenon \cite{Ledbetter:2002}. In addition the sample block must be maintained at a low enough temperature below the $T_c$ of Niobium and lead in order for the superconducting magnetic field to be effective.

Magnetic gradients within the sample block itself can be reduced using superconducting gradient compensation coils, and the applied magnetic field at the \ce{^3 He} sample can be set using a D-shaped half-Helmholtz coil~\cite{Fosbinder-Elkins_2022}.  A shielding factor of order $10^{21}$ near the front surface of the magnetized source mass is required to achieve the sensitivity indicated in Figs.~\ref{fig:sm-dip} and \ref{fig:dip-dip}, which could be achieved through a combination of thin film superconducting shielding and superconducting foils with an appropriate geometry. Acoustic vibrations also need to be attenuated at a similar level required in the ARIADNE experiment~\cite{Arvanitaki:2014dfa}, since they can cause magnetic field variations due to the image magnetization arising from the Meissner effect in the superconducting shields. 

\begin{figure}[h]
\centering
\includegraphics[width=0.99\columnwidth]{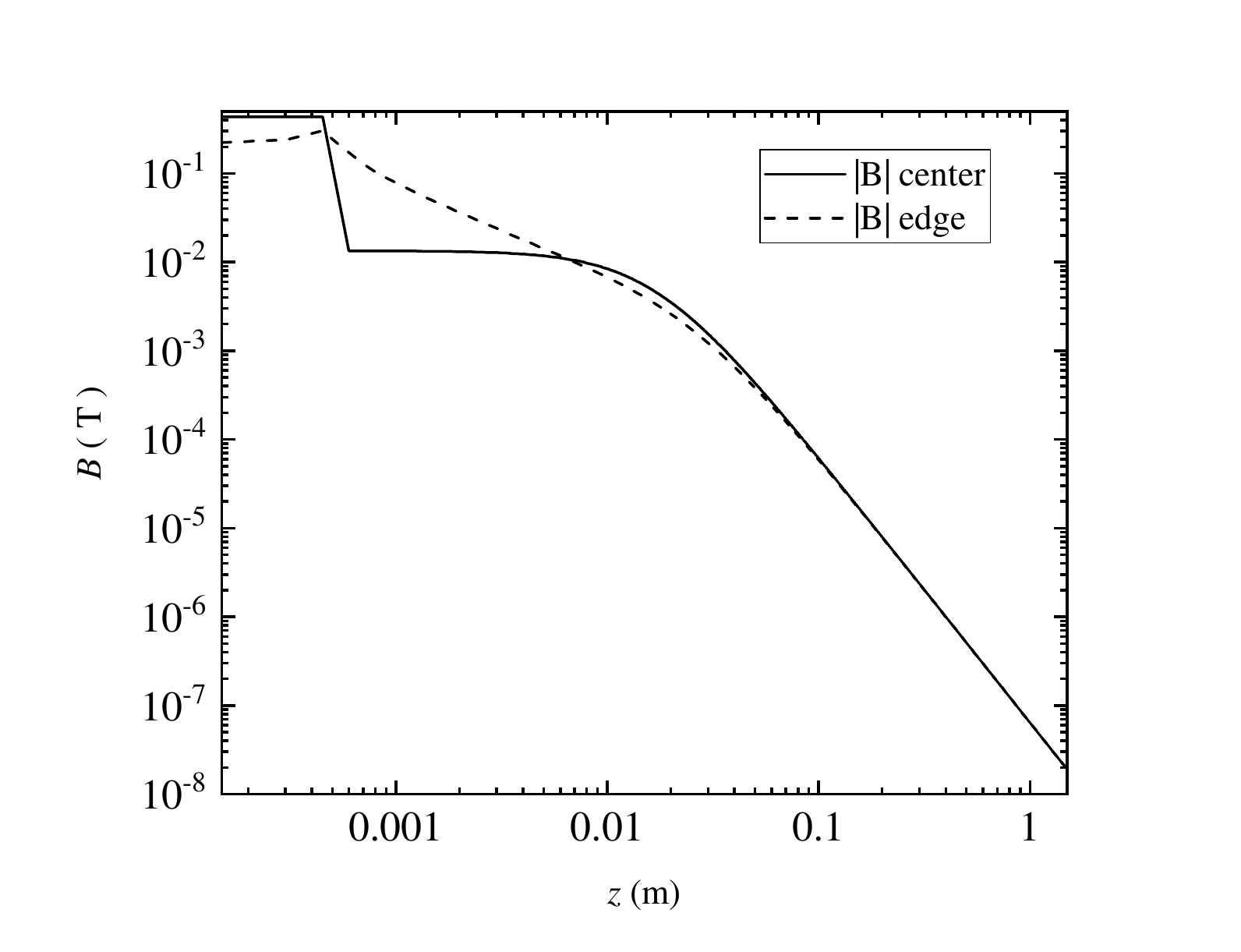}
\caption{Magnetic field magnitude $B$ along the $z$-axis of the sample extending away from the source mass center (edge), shown as a solid (dotted) black line. 
}
\label{fig:Bsource}
\end{figure}

The ferromagnetic source mass will typically have a remnant magnetic field of order $0.1\,\mathrm{T}$, and we estimate the magnetic field near the source crystal to determine the magnetic shielding requirements for the \ce{^3 He} sample chamber in the proposed geometry. Fig.~\ref{fig:Bsource} shows the magnetic field profile extending away from the source mass with $D=1 \, \mathrm{mm}$, assuming a ferromagnetic sample with a magnetization of $3.55 \times 10^5 \, \mathrm{A/m}$ corresponding to the saturation magnetization of Eu$_{0.5}$Ba$_{0.5}$TiO$_3$~\cite{magsource}. The field at the front surface of the sample chamber is estimated to be of order $0.1\,\mathrm{T}$, so the shielding factor required on the front is approximately $10^{21}$.  
We consider
a two-layer superconducting shield, composed of a $1\,\mathrm{\mu m}$ thick Nb film and a $10\,\mathrm{\mu m}$ thick Pb foil. The penetration depth of the foil is approximately $39\,\mathrm{nm}$, which should provide adequate shielding for thicknesses greater than $2\,\mathrm{\mu m}$. In Ref. \cite{Fosbinder-Elkins_2022}, the ARIADNE collaboration experimentally demonstrated in a test setup that a combination of Nb thin film shielding and lead foil attains its required shielding factor of order $10^8$ to achieve spin-projection noise limited sensitivity to magnetic fields in the NMR sample. Although the present requirements are more stringent, we expect that the performance can be enhanced using additional thickness or foil layers.

Although the sample region is enclosed in superconducting shielding, openings are required to facilitate electrical connections to the SQUID readout, as well as the current loops for setting the Larmor frequency at the sample and minimizing the gradient of the field, as described in Ref.~\cite{Fosbinder-Elkins_2022}. To maximally screen the magnetic field of the magnetized source mass from the sample, we consider a single tube opening in the back of the sample chamber. For a tube, the shielding factor depends on the length $l$ and radius $a$ of the tube as $\cosh{(1.84l/a)}$~\cite{Claycomb_1999}. 

It is important to also consider the detail for how the wiring for the SQUID magnetometer and gradient compensation coils and D-shaped Helmholtz coils is connected through the tube opening~\cite{Fosbinder-Elkins_2022}.  If a superconducting wire extends down the length of the shielded tube at the rear of the sample enclosure, due to topological effects of the region not being simply connected, the Meissner screening effect is severely reduced, and the external magnetic field can penetrate into the shielded volume.  To avoid this, a transition from superconducting to normal metal is needed halfway down the length of the tube~\cite{Fosbinder-Elkins_2022}. The effective length of the tube can be conservatively considered as the length of the normal metal portion. From Fig.~\ref{fig:Bsource}, we estimate the magnetic field from the magnetized source mass at the location of the back of the sample block a few centimeters away to be on the order of $10^{-4}\,\mathrm{T}$. We estimate an aspect ratio of $\sim{24}$ and tube length $120 \, \mathrm{mm}$ will suffice to achieve the sensitivity shown in Figs.~\ref{fig:sm-dip}  and~\ref{fig:dip-dip}.

As in the setup used for the ARIADNE experiment~\cite{Arvanitaki:2014dfa}, acoustic vibrations can cause magnetic field variations due to the image magnetization arising from the Meissner effect in the superconducting shields.  For a $0.1\,\mathrm{\mu m}$ wobble in the source crystal stage plate at $\omega_{\rm{m}}/2\pi = 10 \, \mathrm{Hz}$, we roughly estimate a $\delta_x \sim 2 \, \mathrm{nm}$ vibrational amplitude of the sample chamber. The helium gas mixture in our regime has a sound speed of order $300 \, \mathrm{m/s}$, so the motion of the gas molecules can follow the displacement of the shield adiabatically for a motion of this amplitude at $10\,\mathrm{Hz}$.  We therefore expect no significant magnetic field fluctuations provided the quartz enclosure containing the sample moves rigidly. We can estimate the degree to which this approximation breaks down by considering the quartz wall thickness, size, and known elastic properties. 
Assuming $\delta_x =2 \,\mathrm{nm}$, we find that the relative motion between the sample chamber and the shield coating on the outside surface of the quartz block from elastic deformations is $10^{-17} \,  \mathrm{m}$.  With a gradient of $10^{-5} \, \mathrm{T/m}$, this corresponds to a field background of $\sim 10^{-22}\,\mathrm{T}$.  
Incoherent background vibration of the source mass stage plate also should remain below $0.1\,\mathrm{\mu m}$ amplitude at $10\,\mathrm{Hz}$~\cite{geraci2017progress}, since this would produce magnetic field noise of $5 \times 10^{-19} \, \mathrm{T}/\sqrt{\mathrm{Hz}}$ at the resonant frequency, which can begin to limit the sensitivity. For a dry dilution refrigerator, should additional vibration isolation be required, a strategy similar to that used in cryogenic gravitational wave detectors such as KAGRA \cite{KAGRA} and that was also employed in a recent cryogenic dark matter search experiment \cite{Tejas2024} may be possible, where the cold stage can be isolated using geometric anti-spring filters and inverted pendula.

Additionally, thermal noise can lead to vibrational motion in the superconducting film affixed to thin side of the quartz sample block resulting in magnetic field fluctuations. For the quartz wall thickness and dimensions we consider, we estimate that these fluctuations at $4$K can also be kept at or below the level of $5 \times 10^{-19}$ T $/\sqrt{\mathrm{Hz}}$ at the $^3$He nuclear spin Larmor resonance frequency. Such noise can be further suppressed if necessary by slightly increasing the quartz wall chamber thickness or decreasing its area.

\section{MQM Matrix Elements}
\label{app:mqm_elements}
We include here additional details pertaining to the nuclear MQM operator and its matrix elements. For derivations of Eqs.~\ref{eq:C_j} \&~\ref{eq:gamma}, we refer the reader to Refs.~\cite{Khriplovich:1997ga, osti_5511886}. 

The coefficient of Eq.~14 in  the main text is given by:
\begin{align}
\mathcal{C}&=
    \frac{96 (\kappa_1+\kappa_2-2) \text{sgn}(\kappa_1) \text{sgn}(\kappa_2) (\sin  (\pi  (\gamma_1-\gamma_2)))}{\pi {\displaystyle \prod_
    {i=-2}^2\left((\gamma_1-\gamma_2-i)(\gamma_1-\gamma_2+i)  
     \right)}} \label{eq:C_j}\\
     \gamma &= \sqrt{(j+1/2)^2-Z^2\alpha^2}; \quad 
     \kappa = (l-j)(2j-1),\label{eq:gamma}
\end{align}
where $j$ and $l$ are the total and orbital angular momentum quantum numbers, respectively, and the indices $1,2$ refer to the two states in the matrix element of Eq.~14 in the main text. 

We also evaluate the individual electronic matrix elements of 
$\braket{\Omega_{p_{3/2,m_j}}\lvert { \sigma_m \hat{r}_k + \sigma_k \hat{r}_m 
- 2(\bm{\sigma}\cdot \bm{\hat{r}}) \hat{r}_k \hat{r}_m }\rvert \Omega_{s_{1/2, m_{j'}}}}$ 
found in Eq.~14 in the main text. The $\Omega_{jlm}$ are spherical spinor wavefunctions detailed in section 8.1 of Ref.~\cite{Khriplovich:1997ga}. We find the following entries:
\begin{itemize}
\item $\ket{\Omega_{s_{1/2, +1/2}}}, \ket{\Omega_{p_{3/2,+3/2}}}$:
\begin{equation}
    \left(
\begin{array}{ccc}
 0 & 0 & -\frac{\sqrt{\frac{3}{2}}}{5} \\
 0 & 0 & -\frac{1}{5} i \sqrt{\frac{3}{2}} \\
 -\frac{\sqrt{\frac{3}{2}}}{5} & -\frac{1}{5} i \sqrt{\frac{3}{2}} & 0 \\
\end{array}
\right)
\end{equation}
\item $\ket{\Omega_{s_{1/2, +1/2}}}, \ket{\Omega_{p_{3/2,+1/2}}}$:
\begin{equation}
    \left(
\begin{array}{ccc}
 -\frac{\sqrt{2}}{5} & 0 & 0 \\
 0 & -\frac{\sqrt{2}}{5} & 0 \\
 0 & 0 & \frac{2 \sqrt{2}}{5} \\
\end{array}
\right)
\end{equation}
\item $\ket{\Omega_{s_{1/2, +1/2}}}, \ket{\Omega_{p_{3/2,-1/2}}}$:
\begin{equation}
    \left(
\begin{array}{ccc}
 0 & 0 & \frac{3}{5 \sqrt{2}} \\
 0 & 0 & -\frac{3 i}{5 \sqrt{2}} \\
 \frac{3}{5 \sqrt{2}} & -\frac{3 i}{5 \sqrt{2}} & 0 \\
\end{array}
\right)
\end{equation}
\item $\ket{\Omega_{s_{1/2, +1/2}}}, \ket{\Omega_{p_{3/2,-3/2}}}$:
\begin{equation}
    \left(
\begin{array}{ccc}
 \frac{\sqrt{6}}{5} & -\frac{1}{5} \left(i \sqrt{6}\right) & 0 \\
 -\frac{1}{5} \left(i \sqrt{6}\right) & -\frac{\sqrt{6}}{5} & 0 \\
 0 & 0 & 0 \\
\end{array}
\right)
\end{equation}
\item $\ket{\Omega_{s_{1/2, -1/2}}}, \ket{\Omega_{p_{3/2,+3/2}}}$:
\begin{equation}
\left(
\begin{array}{ccc}
 -\frac{\sqrt{6}}{5} & -\frac{1}{5} \left(i \sqrt{6}\right) & 0 \\
 -\frac{1}{5} \left(i \sqrt{6}\right) & \frac{\sqrt{6}}{5} & 0 \\
 0 & 0 & 0 \\
\end{array}
\right)
\end{equation}
\item $\ket{\Omega_{s_{1/2, -1/2}}}, \ket{\Omega_{p_{3/2,+1/2}}}$:
\begin{equation}
\left(
\begin{array}{ccc}
 0 & 0 & \frac{3}{5 \sqrt{2}} \\
 0 & 0 & \frac{3 i}{5 \sqrt{2}} \\
 \frac{3}{5 \sqrt{2}} & \frac{3 i}{5 \sqrt{2}} & 0 \\
\end{array}
\right)
\end{equation}
\item $\ket{\Omega_{s_{1/2, -1/2}}}, \ket{\Omega_{p_{3/2,-1/2}}}$:
\begin{equation}
\left(
\begin{array}{ccc}
 \frac{\sqrt{2}}{5} & 0 & 0 \\
 0 & \frac{\sqrt{2}}{5} & 0 \\
 0 & 0 & -\frac{1}{5} \left(2 \sqrt{2}\right) \\
\end{array}
\right)
\end{equation}
\item $\ket{\Omega_{s_{1/2, -1/2}}}, \ket{\Omega_{p_{3/2,-3/2}}}$:
\begin{equation}
  \left(
\begin{array}{ccc}
 0 & 0 & -\frac{\sqrt{\frac{3}{2}}}{5} \\
 0 & 0 & \frac{1}{5} i \sqrt{\frac{3}{2}} \\
 -\frac{\sqrt{\frac{3}{2}}}{5} & \frac{1}{5} i \sqrt{\frac{3}{2}} & 0 \\
\end{array}
\right)
\end{equation}
\end{itemize}
\end{document}